\documentclass[twocolumn,floats,preprintnumbers,amsmath,amssymb]{revtex4}

\usepackage{graphicx}
\usepackage{dcolumn}
\usepackage{bm}

\begin{document}

\title{Electronic properties of DNA: structural and chemical influence
on the quest for high conductance and charge transfer}
\author{R. G. Endres, D. L. Cox, R. R. P. Singh}
\affiliation{Department of Physics, University of California, Davis, CA 95616}
\date{\today}

\begin{abstract}
Motivated by the wide ranging experimental results on the conductivity of DNA,
we have investigated extraordinary configurations and chemical environments 
in which DNA might become a true molecular wire, perticularly from enhanced
electronic overlaps or from small activation energies.
In particular, we examine A- vs B-DNA, 
the ribbon-like structures proposed to arise from molecular stretching, the 
potential role of counterions in hole doping the DNA orbitals, 
the possibility of backbone conduction, and the effects of water.   
We find that small activation gaps observed in conductivity 
experiments may arise in the presence of water and counter ions.
We further discuss the role of harmonic vibration and twisting motion on 
electron tight binding matrix elements using {\it ab initio} 
density functional theory and model Koster-Slater theory calculations.
We find that partial cancellation between \(pp\sigma\) and \(pp\pi\) 
interaction of \(p_z\) orbitals on adjacent base pairs, along with destructive interference of 
phase factors are needed to explain the weak conductance of A-DNA. 
Our results lead also to a
physical interpretation of the angular dependence of inter-base pair tight binding matrix elements. 
Furthermore, we estimate Franck-Condon factors, reorganization energies 
and nuclear frequencies essential for charge transfer rates, and find
our estimated hole transfer rates between base pairs to be in excellent
agreement with recent picosecond dynamics data.  
   
\end{abstract}

\maketitle

\section{\label{sec:level1}Introduction}
In addition to DNA's fundamental role in genetics, it is 
now also a potential candidate for nano-electronic devices. 
The highly specific binding between single strands of DNA and 
its self-assembly property open a whole new
approach to single-molecule electronics. However, its 
intrinsic conductance properties remain highly controversial. 
As early as 1962, Eley and Spivey suggested that 
\(\pi-\pi\) interactions of stacked base pairs 
in double stranded DNA could lead to conducting
behavior\cite{1962}. Recently, there have been 
several experimental studies of the DNA conductance,
leading to a variety of results. These have ranged from wide-gap insulating behavior
to proximity induced superconductivity.

Although double stranded DNA has similarities to van der Waals stacked 
aromatic conducting crystals ({\it e.g.} the DNA base pair spacing is
similar to the preferred axis lattice spacing of aromatic crystals) 
it also has essential differences from these and conventional conductors. 
Unlike crystals, the DNA is not a periodic system. 
The largest ionization potential difference between two isolated bases
is about 0.6 eV between guanine and thymine which exceeds the estimated
electronic coupling between base pairs.
In addition, DNA's environment,
e.g. its dynamical interaction with 
water and counterions, can have a substantial impact on its conductivity.
From molecular dynamics simulations, the average displacement of a 
base pair in DNA is about \(0.3-0.4 \AA\) \cite{young} 
which is a tenth of the Watson-Crick spacing and an 
order of magnitude higher than in crystals at 
room temperature. All these properties make 
DNA a highly dynamic system and it remains unclear how well traditional 
concepts from solid sate physics can describe it.

DNA has recently been the object of numerous single molecule studies to
ascertain its conducting properties directly.  Results have varied
dramatically, from finding a fully localized insulator\cite{adna}, a
wide gap semiconductor with coherent extended states\cite{porath}, a
metal at room temperature with ohmic current voltage characteristics and
a conductance comparable to doped polyacetylene\cite{fink}, and, most
surprisingly, a proximity effect superconductor at very low
temperatures\cite{proximity}.  Recently, self-assembled networks of DNA have
been developed with quite reasonable conductance properties over
distance scales of 50 nm or less, and high conductance induced by
chemical or gated doping\cite{kawai,fet}; these networks can be used to make
transistor devices. Indeed, a single molecule DNA FET (field-effect transistor) 
has recently been constructed\cite{fet}.
Motivated by these wide ranging experimental results on the conductivity of DNA,
we have embarked on a theoretical effort to ascertain what conditions
might induce such remarkable behavior.  
Our focus here is to examine whether any likely DNA-structures or environments
can yield reduced activation gaps to conduction or enhanced electronic overlaps.
In particular (sec. II), we have studied
a hypothetical stretched ribbon structure, 
A, and B-form DNA, and the effects of counterions (especially sodium and magnesium)
with and without water.  

We have used a combination of fully ab initio density
functional theory (DFT) code (SIESTA\cite{siesta}), and a parameterized 
H\"uckel-Slater-Koster\cite{SK,harris} modeling using only
atomic $p_z$ orbitals ($z$ the long axis of the molecule) of the bases.    

We have been unsuccessful in identifying structural motifs that might
engender metallic conductance in DNA, but we give an explanation why
DNA in the A-form provides less effective electronic coupling along the 
helical axis than B-DNA.
While direct matrix elements between two 
highest occupied molecular orbitals (HOMOs) or 
lowest unoccupied molecular orbitals (LUMOs)
for model A-(B-)DNA are of order 0.01 eV (0.1 eV),
 individual matrix elements between $p_z$ orbitals
on well contacted atoms in adjacent base pairs
can generally exceed 1 eV. 

We also demonstrates the possibility of small activation gaps of order \(k_BT\)
with a first principles study of 4 base pair long B-DNA. We discuss the cases of
sodium and magnesium counterions, with (wet DNA) or without water (dry DNA).

We have also studied the influence of longitudinal and torsional vibrations
upon the electronic structure of DNA (sec. \ref{sec:level3}), and
estimated reorganization energies, Franck-Condon factors, and charge
transfer rates between adjacent bases (sec.\ref{sec:level4}).  
We find good agreement between our estimated rates and recent experimental
data\cite{barton} assuming that torsional (twist) vibrations limit the
charge transfer most significantly. 
Finally, we conclude in section \ref{sec:level5}. Some of the mathematical
details can be found in appendix \ref{sec:levelA}.

\section{\label{sec:level2}Theoretical search for high conductance DNA}

\subsection{\label{subsec:sublevel1}Structural influence}

In this section we demonstrate that the electronic coupling
between base pairs is highly sensitive to the 
DNA-structure and thus to the influence of its environment, which can
modify the structure.

In biological situations, there are
several double helical conformations of DNA depending
on the humidity and salt concentration.
We consider the A- and B-forms.
The B-form predominantly occurs {\it in vivo},
has a unit cell of 10 base pairs (neglecting sequence effects),
a helical rise of about \(3.375 \AA\) per base pair, 
and a twist angle between adjacent base pairs on average
of \(36.0^o\) \cite{Bform}.
Since the base pairs are approximately perpendicular
to the helical axis, the base pair
separation is close to the helical rise. The A-form exists at
low humidity in the presence of some salts. The double
helix is relaxed to 11 base pairs in a unit cell
exposing more of the hydrophobic core
of base pairs and portions of the sugar units of the backbone.
The helical rise and
base pair separation are \(2.56 \AA\) and \(2.425 \AA\),
respectively. The twist angle
is only \(32.7^o\)\cite{Aform}.

\begin{figure}
\includegraphics[width=8.5cm]{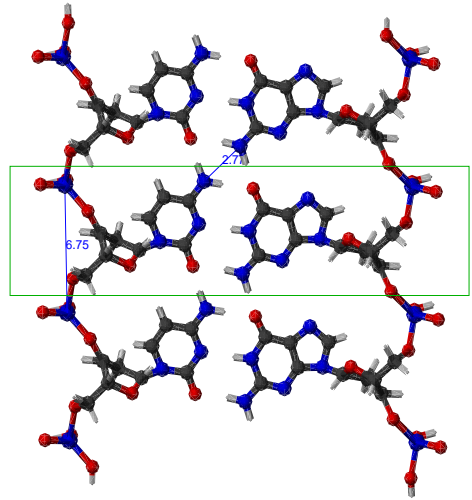}
\caption{\label{fig:fig1}Ideal 2-dimensional ribbon structure for DNA stretched by a factor of 2. The box indicates the unit cell of height \(6.75 \AA\). There are only very few good electronic contacts. }
\end{figure}

There are also other known conformations.
In order to obtain straight DNA molecules
for conductivity measurements,
one often uses a technique called 
molecular combing \cite{comb}. When a DNA molecule
is anchored at one end on a hydrophobic surface like
polystyrene, a receding meniscus can stretch DNA by
a factor of 1.7,
or if grafted at both ends to the surface even
by a factor as large as 2.1.
The capillary forces are typically \(> 160 pN\)
and can become as large as \(500 pN\)
which results in breaking of the molecule.
The stretching has been theoretically examined with
molecular dynamics simulations using classical
force fields \cite{MD}. Since the
phosphate-phosphate distance along the 
backbone is about \(7 \AA\) and the base
pair separation in B-DNA is about \(3.4 \AA\), stretching by
a factor of 2 can, in principle,
lead to a complete unwinding of the double helix.
Since just stretching would
weaken the inter base pair electronic coupling,
we examine the possibility of
base pairs which lie in the plane of the backbone strands
(fig.\ref{fig:fig1}).
In this case, the base pairs can come
close again and the kinetic energy of
the \(\pi\) electrons can in principle
be lowered if there was decent \(pp\pi\)
overlap between the \(p_z\) atomic 
orbitals at the closest contacts
(with the z-axis now perpendicular to the ribbon).
Ribbon-like structures are indeed
found by K. M. Kosikov {\it et al} \cite{MD},
e.g. the structures \(A_1\) or \(B_1\),
by using a classical force field.

\subsection{\label{subsec:sublevel2}{\it Ab initio} estimates of electronic couplings}

In order to get a realistic picture
of the electronic coupling strength between two base pairs,
the smallest structural unit 
through which charge has to go, we use the DFT
code \textsc{SIESTA} \cite{siesta}. Since its aim is
efficiency and large systems, it uses Troullier-Martins norm-conserving
pseudo potentials\cite{ps} in the Kleinman-Bylander form\cite{klein}, 
and a basis set of numerical atomic orbitals using the method
by Sankey and Niklewski\cite{sankey}.
We used a double-\(\zeta\) basis set with polarization orbitals
(DZP), the largest basis within \textsc{SIESTA}, and the
generalized gradient approximation (GGA) for the
exchange-correlation energy functional in the 
version by Perdew, Burke and Ernzerhof\cite{gga}.
The structures for B and A-DNA were obtained
from x-ray diffraction data (\cite{Bform} and \cite{Aform},
respectively).
Since the hydrogen atoms were missing in the A-DNA
structural data, they were added and relaxed with the
conjugate gradient method. The ideal 2-dimensional
ribbon structure of figure \ref{fig:fig1} was created
artificially in order to get most effective \(pp\pi\)
overlap and a small unit cell containing only a single
base pair. As for the 2 base pair calculations,
only methylated\cite{CH3} base pairs without
the backbone were used for simplicity.

Consider two molecular orbitals (MOs), \(\alpha\) from base pair 1, and \(\beta\) from base pair 2.
\(\alpha, \beta\) are eigenstates from SIESTA for individual base pairs. They can be expanded
in terms of the original atomic orbitals of SIESTA, e.g. 
\(|\alpha>=\sum_{i=1}^{N_\alpha}c^\alpha_i|ao>_i\) with \(N_\alpha\) being the dimension of the 
base pair 1, and  \(c^\alpha_i\) and  {\(|ao>_i\)} the LCAO (linear combination of atomic orbitals)
coefficients and atomic orbital basis states, respectively. 
 
The electronic coupling \(t_{\alpha\beta}\) between states \(\alpha\) and \(\beta\) can be computed as follows  
using formulas for non-orthogonal states\cite{newton}, {\it viz}
\begin{eqnarray}
t_{\alpha\beta}&=&[t'_{\alpha\beta}-0.5 (t'_{\alpha\alpha}+t'_{\beta\beta})S_{\alpha\beta}](1-S_{\alpha\beta}^2)^{-1}\label{eq:t12}\\
t'_{\gamma\delta}&=&\sum_{i=1}^{N_\gamma}\sum_{j=1}^{N_\delta}c_i^\gamma c_j^\delta H_{ij}^{\gamma\delta},\quad S_{\alpha\beta}=\sum_{i=1}^{N_\alpha}\sum_{j=1}^{N_\beta}c_i^\alpha c_j^\beta S_{ij}^{\alpha\beta},
\end{eqnarray}
where \(\gamma\) and \(\delta\) can take the values \(\alpha\) or \(\beta\) 
according to eq. \ref{eq:t12}.
The total Hamiltonian H (overlap S) matrix of
the two base pair system with dimensions \((N_\alpha+N_\beta) \times (N_\alpha+N_\beta)\) 
can be obtained from SIESTA. It can be thought of as 
consisting of two diagonal blocks \(H^{\alpha\alpha}\), \(H^{\beta\beta}\) 
(\(S^{\alpha\alpha}\), \(S^{\beta\beta}\))
representing the intra-base pair couplings (overlaps)
and one off-diagonal block \(H^{\alpha\beta}\) 
(\(S^{\alpha\beta}\)) representing inter-base pair couplings (overlaps).

The values for the electronic couplings \(t_{\alpha\beta}\) 
between two HOMOs, as well as two LUMOs, 
are shown in table \ref{tab:table1}. 
As for the B-form, they are in general small and in good agreement with earlier
{\it ab initio} studies \cite{band_IP,band,elcoupl}, but
for A and ribbon form they are much smaller than for the
B-form. For the ribbon structure (including the backbone), 
we also carried out a band structure calculation after fully relaxing the 
geometry. The bands were calculated at 30 k points
in the long direction using a Brillouin zone sampling at 3 k points.
There was no sign of dispersion in agreement
with the weak electronic coupling. That is, once stretched the base pair
states localize.

\begin{table}
\caption{\label{tab:table1}Transfer integrals in eV between HOMOs (\(t_H\)) and LUMOs (\(t_L\)) of two 
base pairs. The calculations are done with \textsc{SIESTA} and DZP basis set. The expression dimer 
always stands for a two base pair long DNA sequence, e.g. the GG dimer stands for 5'-GG-3' DNA.}
\begin{ruledtabular}
\begin{tabular}{ccccc}
           &\multicolumn{2}{c}{A-DNA} &\multicolumn{2}{c}{B-DNA} \\ \hline
dimer      & \(t_H\)   & \(t_L\)      & \(t_H\)   & \(t_L\)      \\ \hline
GG         & 0.0069   & -0.0006      & -0.1409   & 0.0525      \\ \hline  
GG\((2.425 \AA)\)(ref. \cite{f1})
           &           &              & -0.6922  & 0.2548     \\ \hline 
GG\((0.0^o)\)(ref. \cite{f2})
           &           &              & 0.2385   & 0.3233     \\ \hline
5'-AG-3'   & -0.0153  &  0.0060     & -0.0710  & 0.1124     \\ \hline
5'-GA-3'   & -0.0113  & -0.0010     & -0.1871  & 0.0472     \\ \hline
AA         & 0.0310   & -0.0120     & -0.0695  & 0.1054     \\ \hline       
           &\multicolumn{2}{c}{Ribbon}&           &              \\ \hline
GG         & 0.0039   & 0.0083      &           &              \\                        
\end{tabular}
\end{ruledtabular}
\end{table}

Thus, our conclusion is that
dehydrating (A conformation) and stretching of
DNA should lead to even more highly localized states and 
insulating behavior than in B-DNA. For the ribbon, the
main reason is that there are only very few (\(\approx 3\))
close contacts (\(\approx 2.7 \AA\)) between neighboring base pairs
(see also figure \ref{fig:fig1}). For A-DNA, the weak
coupling might seem a bit
surprising since the base pair separation is much
smaller than in B-DNA, but it can be
understood in terms of less effective stacking. This is explored 
in the next subsection.

\subsection{\label{subsec:sublevel3}Slater-Koster modeling of coupling strength}

\begin{figure}
\includegraphics[width=6.5cm]{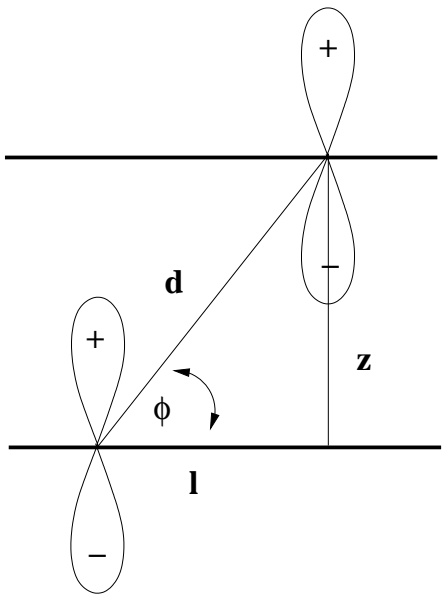}
\caption{\label{fig:fig2}Inter-atomic matrix elements from Slater-Koster theory: two atomic \(p_z\) orbitals on adjacent base pairs couple by positive pp\(\sigma\) and negative pp\(\pi\) interactions.}
\end{figure}

In order to understand the results in the previous subsection
more quantitatively, we have performed model calculations 
varying the angular amd spatial separation of bases.
These are more readily performed than full {\it ab initio} calculations
and often provide more physical insight. 
We restrict ourself to just atomic \(p_z\) orbitals, where the 
\(z\) axis can be defined locally by the normal of the corresponding base.
In this case, two \(p_z\) orbitals from different base pairs 
couple by pp\(\sigma\) and pp\(\pi\) interactions, which we model with semi-empirical
Slater-Koster theory\cite{SK,harris}
\begin{equation}
V_{ppX}=\eta_{ppX}\frac{\hbar^2}{md_o^2}e^{-d/R_c}\label{eq:ppX}.
\end{equation}
We note \(\eta_{pp\sigma}>0\) and  \(\eta_{pp\pi}<0\). 
Here, \(d\) and \(m\) are 
distance and electron mass, and \((\hbar^2)/(md_o^2)=7.62 eV\).
The exponential distance cut-off \(R_c\) and \(\eta\) are 
parameters to be determined by matching to {\it ab initio}
results as discussed below. The interatomic
matrix element between two ``parallel'' \(p_z\) orbitals depicted in figure \ref{fig:fig2} is 
\begin{eqnarray}
E_{zz}&=&\sin^2\phi V_{pp\sigma} + \cos^2 \phi V_{pp\pi}\label{eq:V}\\
&=&\frac{\hbar^2e^{-d/R_c}}{md_o^2}\left[(\eta_{pp\sigma} + |\eta_{pp\pi}|)\frac{z^2}{l^2+z^2}-|\eta_{pp\pi}|\right]\nonumber,
\end{eqnarray}
According to this formula, the combination of close base pair separation \textit{and}
poor contacts reduces the electronic coupling between base pairs, i.e. 
can lead to cancellation effects.

A general formula for non-parallel orbitals, 
which is used for all our following model calculations 
can be found in the Appendix, eq. (\ref{eq:VSK}). 
The normal of each base is taken for the direction for 
all \(p_z\) orbitals within that base, and is obtained by  
fitting a plane to each single base whose coordinates
were obtained from x-ray diffraction data for A and B-DNA.
This formula constitutes a straightforward way to calculate
inter - base pair couplings for a variety of spatial and 
angular separations.

The Slater-Koster parameters \(\eta_{pp\sigma}\),
\(\eta_{pp\pi}\) and cut-off \(R_c\) were determined
from fitting to DFT data as follows: 
Starting from the DZP basis of SIESTA,
we first identify the \(2p_z\) orbitals. These are
the first-\(\zeta\) orbitals of \(p_z\)
symmetry. (The second-\(\zeta\) orbitals
correspond to excited atomic states and have rather small 
contributions to the wavefunctions and matrix elements [\(\lesssim 20\%\)].) 
We consider the reduced SIESTA Hamiltonian matrix
in this basis, which has two diagonal blocks
representing the on-site and intra-base pair couplings and
one off-diagonal block representing
inter-base pair couplings. The Slater-Koster matrix elements
calculated with eq. (\ref{eq:VSK}) are
directly fitted to the off-diagonal block matrix elements. 
The fitting was done with simulated annealing, which gave
better (and still physical)
results than e.g. with the Powell algorithm\cite{fit}. Finally, 
since the DFT code uses a global z-axis along the helix
which is different from the local z-axis due to
an inclination and a propeller twist angle associated
with each base, we turn to parallel reference base pairs
for the fitting process.
In the case of A-DNA we use two parallel \(G\cdot C\) base 
pairs with a separation of \(2.43 \AA\),
while for B-DNA ones with a separation of \(3.34 \AA\).
The fitted parameters are \(\eta_{pp\sigma}=2.93\), 
\(\eta_{pp\pi}=-0.73\), \(R_c=1.16\AA\) for A-DNA, and  
\(\eta_{pp\sigma}=5.27\), \(\eta_{pp\pi}=-2.26\),
\(R_c=0.87\AA\) for B-DNA.

\begin{figure}
\includegraphics[height=8.5cm,angle=-90]{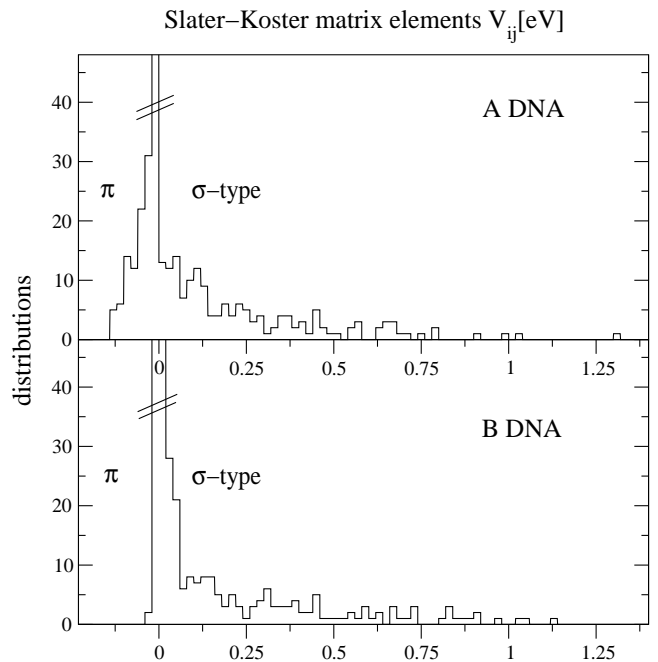}
\caption{\label{fig:fig3}Distribution of Slater-Koster inter-atomic matrix elements between \(p_z\) orbitals of two stacked  $G\cdot C$  base pairs. There are a total of 361 matrix elements. The smallest peak height corresponds to a single matrix element. Details about the distributions, mean: A 0.09, B 0.13; standard deviation: A 0.21, B 0.23; number of matrix elements above 0.75eV: A 6, B 13.}
\end{figure}

In order to further stress the difference between A and B-DNA, 
figure \ref{fig:fig3} shows a distribution
of all possible \(p_z\) interatomic matrix elements
between two \(G\cdot C\) base pairs in the A-form and in the B-form.
Since guanine and cytosine have 11 and 8 \(p_z\) orbitals, respectively
there are a total of \(19\times19=361\) matrix elements.
(Adenine and thymine have 10 and 8 \(p_z\) orbitals respectively.)
Figure \ref{fig:fig3} shows that there are more good
contacts, e.g. twice as many above 0.75 eV, in the B-form than in A-form.
Furthermore, there is a shift to (negative) \(pp\pi\) interaction in 
the A-form, although the single largest matrix element 
(\(\approx 1.3\) eV) arises in the A-form. This is presumably due
to a single optimal contact at the shorter distance.

Table \ref{tab:table1} shows the electronic couplings between 
specific MOs which correspond to HOMOs and LUMOs. A way to understand the smallness
of \(t_{HOMO}\) (\(t_H\)) and \(t_{LUMO}\) (\(t_L\)) for A-DNA relative to B-DNA
is to note that the distribution (fig. \ref{fig:fig3}) is clearly more 
centered around zero in case of A-DNA, while for B-DNA it is more asymmetric with
mostly positive matrix elements.  The electronic coupling is a weighted average 
of these matrix elements  with the weights being the product of two LCAO coefficients.
It can be expected that the difference in the distributions leads to small
electronic couplings for A-DNA.

Since A-DNA appears clearly insulating, as already indicated by the small bandwidth 
reported in ref. \cite{adna}, we restrict ourselves to B-DNA in the subsequent sections.

\subsection{\label{subsec:sublevel4}Effect of counterions ions and water on electronic structure}

DNA is a highly charged molecule which can only be stable if the 
negatively charged phosphate groups are
neutralized by positive counterions or by polarized water molecules 
from a buffer solution. In order to measure its conductivity 
DNA has initially to be dried, normally by a flowing \(N_2\) gas\cite{kawai}. 
This can result in as few as 2 to 3 water molecules per nucleotide\cite{tran},
which is not even high enough humidity to obtain the A-structure. In principle dried DNA
can turn into a disordered structure\cite{hopfinger} 
probably not suitable for
charge transport through the base pairs at all. Evidence for that is
found by STM images of dried DNA on mica,
which shows a base pair separation of \(\sim 7.2 \AA\) \cite{Cai2}, 
and also by AFM images on \(SiO_2\) which show a DNA sample height
of only 0.5 nm instead of the diameter 2nm of the usual double helix\cite{storm}.

Different research groups in general work with 
different buffer solutions: activated 
(0.12 - 0.2 eV) hopping conductivity with an initial sodium ion
buffer was observed by Kawai's group \cite{fet}
using \(\mu\)m long poly(G)-poly(C)/poly(A)-poly(T) DNA, and 
by Rakitin {\it et al} \cite{rakitin}, as well as Tran 
{\it et al} \cite{tran} using \(\lambda\)-DNA. 
On the other hand, proximity effect superconductivity was observed by
Kasumov {\it et al} \cite{proximity}, who used 
a few 16-\(\mu\)m-long \(\lambda\)-DNA molecules taken from a 
magnesium buffer solution. 
                               
Motivated by this, we have examined the effects
of various counterions (protons, Na, Mg) and water
on the electronic orbitals of DNA. 
We use a 4 base pair long 5'-GAAT-3' structure 
obtained from classical molecular dynamics (MD) with the AMBER 4.1 force field.
The structure was taken from a snap shot
of the Dickerson dodecamer with thousands of water molecules \cite{young}
and was provided to us by S. Dixit. 
For wet DNA we keep the 
first and second solvation shell (\(\sim\) 25 water molecules per nucleotide),
which corresponds to keeping all the water molecules within a 
4.5 \(\AA\) radius of the DNA atoms. Although B (wet) DNA exists
only above 13 to 18 water molecules per nucleotide, we also examine completely
dry DNA by removing all the water molecules but keeping
the same B-DNA-structure. From our own experience with DFT calculations,
DNA orbitals are generally rather localized
on various groups like phosphate, sugar, counterions, and bases. 
Thus it might still be possible that this artificial dry DNA 
shows the correct order of energy levels and hence gives
valuable information about activation gaps. 
The location of the counterions are kept the same as for
wet DNA, i.e. some are further outside the DNA helix while others are at the
phosphates or in the grooves. 
Counterion distributions are discussed in ref. \cite{ions}. 
For comparison we also considered protonated DNA of the same structure, since
this is what theorists usually use for neutralized DNA. This has the 
big advantage of keeping the number of atoms small. 
The proton counterions are placed at the phosphates where they 
bind covalently. All other metal counterions and water molecules are removed. 

We optimized the geometries with a few conjugate gradient steps (time consuming), 
but the forces still large (of order eV/\(\AA\)). On can argue that this 
out-of-equilibrium structure might nevertheless resemble DNA 
in a solution at finite temperature where forces are always non-zero due
to the motion of atoms and molecules. 
As a further check for dry DNA, we placed the metal ions at different
locations and observed no principle changes occurred
in the band structure or the Mulliken population\cite{mulliken} of the ions.
For DNA we use a double-\(\zeta\) basis set except for phosphorus, hydrogen
involved in hydrogen bonding and the counterions for which we also 
include polarization orbitals. For the water molecules (more than 200) 
we only use a minimal, single-\(\zeta\) (SZ) basis. The density of states (PDOS) 
is then projected onto atomic orbitals for energies around 
the Fermi energy \(\epsilon_F\) \cite{Fermi} to see which atomic orbitals 
can contribute to charge transport and what effects the counterions have on
the eigenvalue spectrum.

The PDOS of the ith atomic basis orbital projected on the nth
MO is
\begin{equation}
\rho_i^n(\epsilon)=\sum_j^{N_B}c_i^n\ c_j^n\ S_{ij}\ \delta(\epsilon-\epsilon_n)\label{eq:pdos},
\end{equation} 
where \(N_B\) is the dimension of the basis set, \(c_i^n, c_j^n\)
are the LCAO coefficients and \(S_{ij}\)
are the atomic overlap matrix elements. We project on the atomic \(p_z\) orbitals
of the elements C, N and O of all the bases (resulting in \(\pi\) and \(\pi^*\)
MOs), all orbitals of the five phosphate atoms, of the sugar atoms, 
counterions, and also whole water molecules.

\begin{figure}
\includegraphics[height=8.5cm,angle=-90]{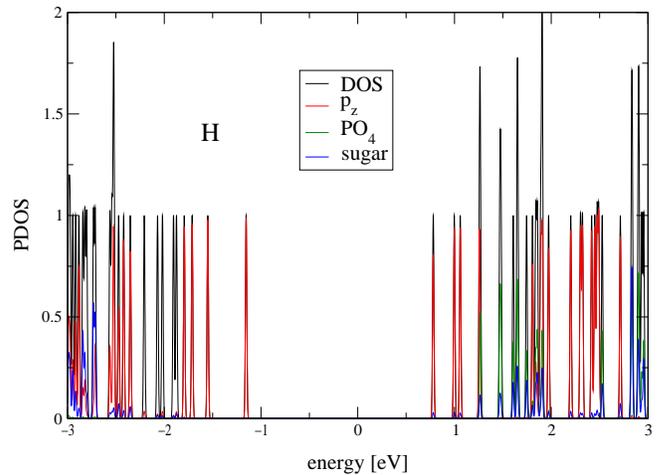}
\caption{\label{fig:fig4} PDOS of protonated 4 base pair long B-DNA; black: total DOS, red: \(p_z\) orbitals of atoms
      in base pairs, green: phosphate, blue: sugar. The Fermi energy is set to 0 eV.}
\end{figure}

The result for the protonated DNA is shown in figure \ref{fig:fig4}.
The atomic \(p_z\) orbitals (red) of the bases form an 
expected \(\pi\) - \(\pi^*\) gap of about 1.9 eV around 
\(\epsilon_F\) (set to zero). One has to keep in mind that DFT 
usually underestimates the gap in insulators by about 50 to 100\%.
The HOMO, mainly guanine, is about 0.4 eV
higher than the next lower occupied 
MOs which are mostly adenine.
The LUMO (cytosine) is about 0.2 eV lower than the next higher unoccupied
MO (thymine). The occupied (unoccupied) phosphate states (green)
and sugar states (blue) on the other hand are energetically lower (higher) than 
the \(\pi\) (\(\pi^*\)) orbitals.

\begin{figure}
\includegraphics[height=8.5cm,angle=-90]{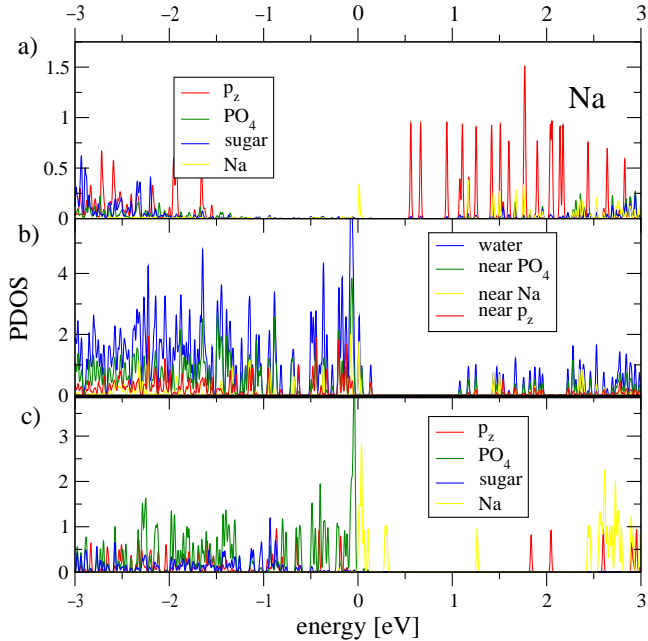}
\caption{\label{fig:fig5}Effects of sodium counterions and water on the molecular orbitals of 4 base pair long B-DNA.
      Part a) and b) show the PDOS of wet DNA; a) red: \(p_z\), green: phosphate, blue: sugar, 
      yellow: sodium; b) blue: all water molecules, green: water molecules near (\(< 3.5 \AA\)) phosphates, 
      yellow: water molecules near sodium, red: water molecules near bases. In part c) all the water are removed (dry DNA).  }
\end{figure}

The results for sodium are shown in figure \ref{fig:fig5}. Parts a) and b)
contain the PDOS of wet DNA, i.e. part a) shows the projection onto
the bases (\(p_z\)), phosphates, sugars, and sodium orbitals, while b)
shows the projection onto water. In a) one can see a \(\pi\) - \(\pi^*\) gap 
of about 2.2 eV. It has slightly increased in the presence of water 
compared to protonated DNA. 
The highest \(\pi\) states around -1.6 eV (relative to \(\epsilon_F\)) 
are mainly guanine,
the next lower group of occupied states (\(\sim -1.9\) eV) are
quite delocalized on several adenines with some guanine admixture. 
The first \(\pi^*\) state has mainly thymine character, the next higher also
a little bit of cytosine. 
In the presence of water the \(\pi\) and \(\pi^*\) MOs seem 
to be more extended than in case of protonated DNA,
where the MOs are more localized on individual bases.

Although the \(\pi\) - \(\pi^*\) gap is rather large, the real HOMO - LUMO gap
is surprisingly small (54 meV). Electrons could possibly be excited
from water levels below \(\epsilon_F\) into unoccupied water and sodium states.
The presence of water and counterions might result in a weak conductance 
if a voltage is applied. 
Part b) of figure \ref{fig:fig5} separates the water 
PDOS (blue) into contributions coming from water molecules near the phosphates (green), 
near sodium (yellow), and near the bases (red). As one can see, the water states 
around \(\epsilon_F\) are mostly from water molecules near the sodium ions. 
The LUMO is a mixture of water and sodium. Furthermore,
the negatively charged phosphates and positively charged sodiums are affecting
the electronic structure of water drastically, since the several-eV large insulating
gap of water is reduced to about 1 eV (0.7 eV in case of magnesium). 
A Mulliken population analysis of sodium 
reveals that sodium is on average a \(+0.88\) charged ion. 
This is consistent with first principle
studies of solvated metal ions, which indicate that solvated
sodium is nearly a \(+1\)-ion almost independent of the water coordination number\cite{solv_ions}. 
Regarding the sodium distribution, two sodiums are several 
\(\AA\) outside the DNA double helix, but the Mulliken population 
is hardly any different for these.

In order to see the effects of water on DNA more drastically, part c) of 
figure \ref{fig:fig5} shows the DOS of DNA with all the water molecules removed. Since the
sodium ions (now only +0.66) are not directly located at the phosphates, they 
cannot screen their negative charges. Repulsive interactions likely between
the oxygens of each phosphate group increase their energy (green) even beyond the 
\(\pi\) states (red). Although the \(\pi\) - \(\pi^*\) gap is about 2 eV, there
are sodium states only 39 meV above the HOMO (phosphate). 
The PDOS is remarkably reproducible
for different counterion configurations.
One can speculate that due to the small electron excitation gap
hole hopping through the backbone might be possible. Even if the B-DNA-structure 
is unstable under perfectly dry conditions, the 
quasi one-dimensional pathway for conduction through the backbone might still survive.
One the other hand, the insulating sugars (blue) 
between the phosphates do not contribute much to the relevant states and
constitute tunneling barriers.

\begin{figure}
\includegraphics[height=8.5cm,angle=-90]{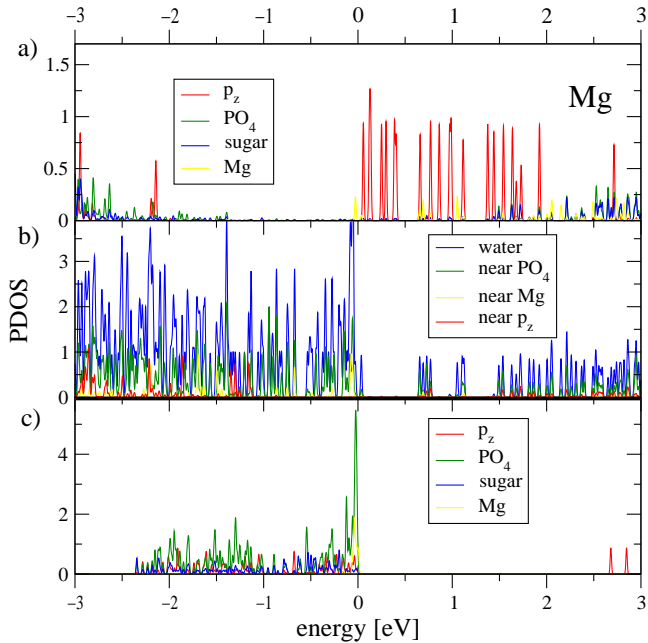}
\caption{\label{fig:fig6}Same as figure \ref{fig:fig5} except for magnesium counterions instead of sodium. Part c): there
      is a single magnesium state right above the Fermi energy (\(\epsilon_F=0\)).}
\end{figure}

For magnesium counterions, 
the result is shown in figure \ref{fig:fig6}. Parts a) and b)
contain the PDOS of wet DNA. Part a) shows the projection on
DNA, while b) shows the projection onto water. 
In a) one can see a \(\pi\) - \(\pi^*\) gap 
of about 2.2 eV which is essentially the same as in the case of sodium. 
The highest \(\pi\) states at about -2.2 eV are a mixture of guanine and adenines,
and 0.8 eV above an occupied adenine/thymine MO.  
The \(\pi\) - \(\pi^*\) gap is again much bigger than the real HOMO - LUMO gap,
which is only 62 meV. Magnesium and water states
below \(\epsilon_F\) could function as electron donors with electrons 
being excited into unoccupied \(\pi^*\) MOs leading to a possible  
electron conduction mechanism.
From part b) of figure \ref{fig:fig6} one realizes however,  
that the water states right below \(\epsilon_F\) 
are not from water molecules in close proximity to the bases 
making the electron donation harder.
The occupied water states from water near the bases are about 1 eV below the \(\pi^*\) MOs. 
A Mulliken population analysis of magnesium 
reveals that magnesium is not a \(+2\)-ion, but has a reduced \(+1.25\) charge. 
One magnesium is more distant from the DNA molecule and is completely surrounded 
by water (7 water molecules within a radius of $3\AA$). 
It has even a smaller (\(+1.0\)) charge. The smaller charge (less than +2) is again 
consistent with first principle studies of solvated metal ions, 
which indicate that magnesium solvated by 4 water molecules has a 
\(+1.1\) charge, but less solvated magnesium has a larger positive charge
closer to \(+1.7\) \cite{solv_ions}.

Part c) of figure \ref{fig:fig6} shows the DOS of DNA 
with all the water molecules removed. Although the 
\(\pi\) - \(\pi^*\) gap is about 2.8 eV, there
is a single magnesium state (not from the further outside one) 
only 11 meV above the HOMO (phosphate). The PDOS is again 
remarkably reproducible for different counterion configurations.

Finally, we try to give a possible explanation for the accumulation or pinning  
of phosphate states right below the metal states for dry DNA. Left unscreened, 
the occupied phosphate states will rise in energy due to repulsive interactions
between negatively charged oxygens of a phosphate group. 
However, they cannot get higher than all the empty metal states. This
is because if they were higher, the phosphates would dump electrons in the metal states 
and reduce their repulsive interactions. This would lower their energy, but if below 
the metal states, the phosphates would accept the electrons again, 
and the same process would repeat itself leading to the pinning-effect.

\section{\label{sec:level3}Dynamical influence on electronic structure}

At finite temperature the base pairs of DNA will
oscillate about their equilibrium positions.
Within our model from section \ref{subsec:sublevel3}
we are ultimately interested in the temperature dependence
of the \(\pi\) - \(\pi^*\) band gap
\begin{eqnarray}
  \Delta_{\text{DNA}}(T)&\approx &\Delta_{\text{single bp}}-\nonumber \\
&&\!\!\!\!\!\!\!\!2(<\!\!t_{\text{HOMO}}\!\!>(T)+<\!\!t_{\text{LUMO}}\!\!>(T))\label{gap},
\end{eqnarray}
with \(\Delta_{\text{single bp}}\) being the temperature independent 
HOMO-LUMO gap of an isolated base pair, 
and \(<\!\!t\!\!>(T)\) describing the temperature dependent average 
electronic coupling between adjacent base pairs. 
(In this model calculation, the \(\pi\) - \(\pi^*\) gap corresponds to the 
HOMO-LUMO gap, because we do not consider water here.) 
The goal is to understand the
experiment by Porath {\it et al} who measured
a strong increase of the voltage gap \(U_c\) (the minimal bias voltage 
above which conduction sets in) with temperature (2-4 eV per 300 K)\cite{porath}. 
In the experiment, short (30 base pair long) homogeneous poly(G)-poly(C)
DNA molecules showed band-like semi-conducting behavior.
In our calculation, we assume that displacements along the helical axis 
and the twisting motion are the most important degrees of freedom which
affect the electronic coupling.
For small changes in the variables, one can assume that these two
degrees of freedom are independent.

Let \(\phi\) and \(u\) describe the deviations from
equilibrium twist angle and base pair separation.
Within our Koster-Slater-model in the CNDO 
(complete neglect of differential overlap) 
approximation, the transfer integral between
two MOs of adjacent base pairs
is given by eq. \ref{eq:t12} with \(S_{12}=0\). 
(From our first principles calculations, \(0.001\lesssim S_{12}\lesssim 0.01\).)
Thus we have
\(t_{12}=\sum_{i,j}\tilde H^{12}_{ij}c^1_ic^2_j\) where the LCAO coefficients
are summed over each of the two base pairs and kept
fixed for all \(\phi\) and \(u\). They are given by
the first-\(\zeta\) orbitals of
\(p_z\) symmetry from a single base pair DZP calculation, and
are corrected by a factor
\(\cos^{-1}\theta\) where \(\theta=8.0^o\) is the angle
between the local base normal and the helical axis.
As noted earlier, the second-\(\zeta\) orbitals yield \(\lesssim 20\%\) 
corrections to this.
The matrix \(\tilde H^{12}\) is obtained as described in 
sec. \ref{subsec:sublevel3} and depends on \(\phi\) and \(u\) through the 
Slater-Koster form of the matrix elements. The tilde indicates that the reduced 
(first-\(\zeta\)) basis set was used.
Since we calculate transfer integrals in
CNDO, we do not need on-site energy matrices \(\tilde H^{11}\), \(\tilde H^{22}\) here.
Nevertheless, these can be easily obtained in the same
fashion. The resulting model can be used
for very large scale quasi-self-consistent-field quantum
calculations of long DNA sequences - similar
to the one used for electron transfer rate calculations
between various donors and acceptors in DNA\cite{huckel}.

\begin{figure}
\includegraphics[height=8.5cm,angle=-90]{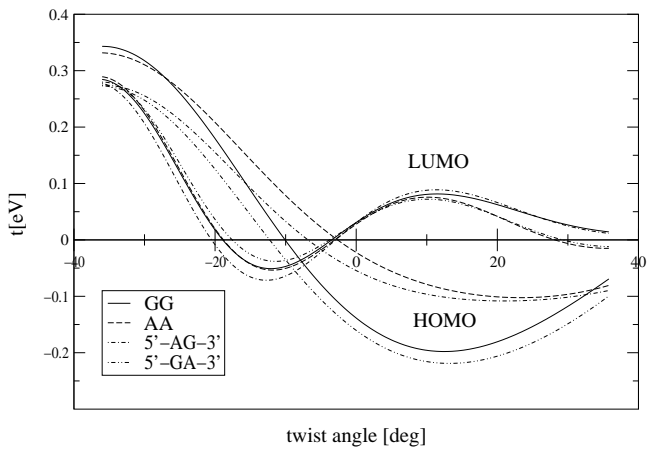}
\caption{\label{fig:fig7}Change of electronic coupling between frontier orbitals of two base pairs
 with relative twist angle \(\phi\) about the equilibrium position. The base pair 
 separation is kept fixed at \(z=3.375\AA\).}
\end{figure}

\begin{figure}
\includegraphics[height=8.5cm,angle=-90]{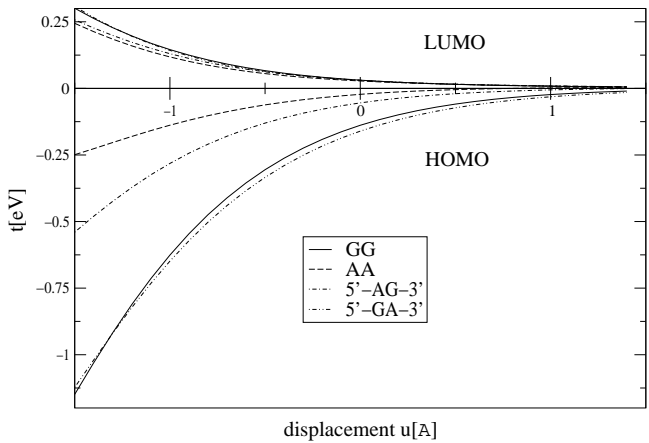}
\caption{\label{fig:fig8}Change of electronic coupling with relative displacement \(u\) about the equilibrium position.
 The twist angle is kept fixed at \(36.0^o\).}
\end{figure}

In figures \ref{fig:fig7} and \ref{fig:fig8} we show
the electronic coupling between the frontier orbitals
as a function the variables \(\phi\) and \(u\). One of them is held
fixed at the equilibrium value, while the other one is varied.
Consider first the variation with the angle \(\phi\) (fig. \ref{fig:fig7}).
At \(\phi=-36^o\), i.e. when the two base pairs are aligned perfectly parallel,
the coupling is maximal and positive. 
Let us focus on a GG dimer. At \(\phi=-36^o\), there are  
19 optimal contacts right on top of each other, and the transfer integral is
approximately
given by \(t\sim\sum_i\tilde H^{12}_{ii}(c_i)^2\) (\(c_i:=c^1_i=c^2_i\)) 
where \(\tilde H^{12}_{ii}>0\) is solely due to \(pp\sigma\) interaction.
Note the interesting property that there are sign changes
of \(t(\phi)\) with \(\phi\). These provide a direct 
explanation for the local minima of \(|t|\) observed in earlier {\it ab initio} studies
\cite{band_IP,band}. In B-DNA, essentially all the interaction
matrix elements (\(\tilde H^{12}_{ij}\gtrsim 0\)) 
are positive (\(pp\sigma\) dominated), 
as can be seen in figure \ref{fig:fig3} 
and this is true for arbitrary twist angles \(\phi\). Since
for larger twist angles, i.e. \(\phi\approx0\),
the transfer integral is 
approximately given by \(t\sim\sum_{i\not=j}\tilde H^{12}_{ij}c^1_ic^2_j\),
the sign change of t can only come from the signs
of the LCAO coefficients (phases of the \(p_z\) orbitals).
This can lead to positive or negative t,
or to a complete cancellation.

\begin{figure}
\includegraphics[height=8.5cm,angle=-90]{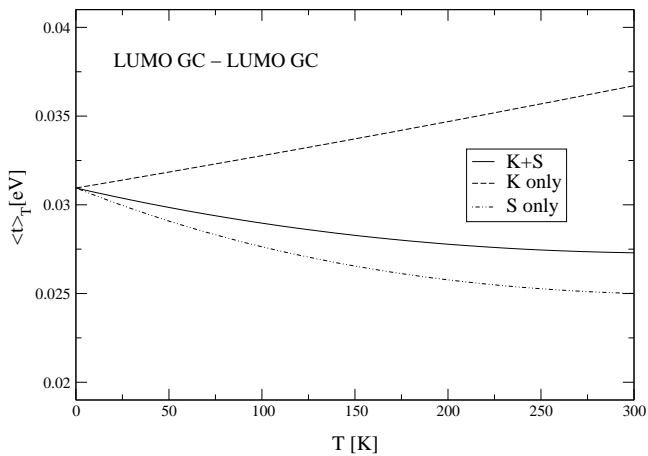}
\caption{\label{fig:fig9}Boltzmann averaged electronic coupling between LUMOs of two $G\cdot C$  base pairs as a function of temperature T. The effects of a spring constant K and a shear constant S are shown.}
\end{figure}

\begin{figure}
\includegraphics[height=8.5cm,angle=-90]{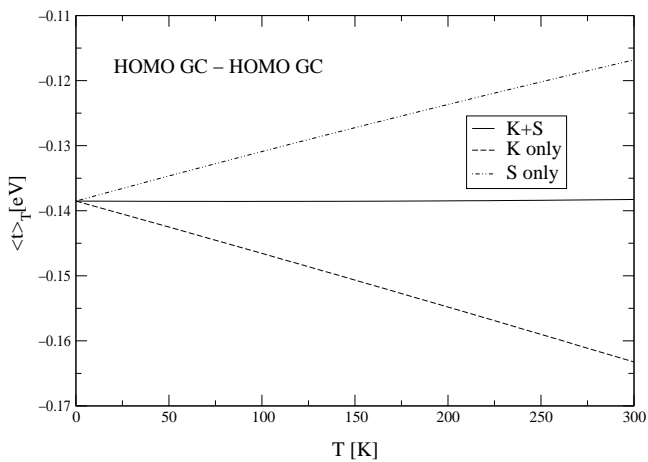}
\caption{\label{fig:fig10}Boltzmann averaged electronic coupling between HOMOs of two $G\cdot C$  base pairs as a function of temperature T. The effects of a spring constant K and a shear constant S are shown.}
\end{figure}

Coming back to the temperature dependence of t and the band gap, we assume 
classical harmonic potentials with spring constant K and shear
constant S for each couple of base pairs. They can be
estimated using the equi-partition theorem from classical
statistical mechanics and estimates for the standard
deviation from molecular dynamics\cite{young},
\begin{eqnarray}
0.375^2\AA^2&=&<u^2>=\frac{k_BT}{K}\label{eq:MDnK}\\
\longrightarrow K&=&0.184 \frac{eV}{\AA^2}=294.8\frac{pN}{\AA}\label{eq:K}\\
{7.5}^2{\text{deg}}^2&=&<\phi^2>=\frac{k_BT}{S}\label{eq:MDnS}\\
\longrightarrow S&=&4.6\cdot 10^{-4}\frac{eV}{{\text{deg}}^2}.\label{eq:S}
\end{eqnarray}
With these estimates for K and S the average electronic coupling
can be calculated
\begin{equation}
<t>(T)=\frac{\int du d\phi\ t(u,\phi)\ e^{-\frac{\beta}{2}(Ku^2+S\phi^2)}}{\int du d\phi\ e^{-\frac{\beta}{2}(Ku^2+S\phi^2)}},\label{eq:t_T}
\end{equation}
with \(\beta=1/k_BT\). The transfer integral only depends on the
differences \(u=u_2-u_1\) and \(\phi=\phi_2-\phi_1\), where
\(u_i\) and \(\phi_i\) are the displacement and twist angle of
the ith base pair. The results are shown in figures \ref{fig:fig9}
and \ref{fig:fig10}. Two things
are remarkable about this result. First, the temperature
dependence is very weak (a few meV change over hundreds of K).
Second, harmonic displacement along the helical axis alone can
lead to an increase
in magnitude of the electronic coupling 
with increasing temperature T, while
twisting motion on the other hand has the opposite effect
leading to a cancellation.

This can be understood by looking at figures \ref{fig:fig7}
and \ref{fig:fig8}.
Since \(t(\phi=0)\) is closer to a local maximum (at
least for HOMO states), a broadening of the
gaussian distribution \(\sim e^{-\beta S\phi^2/2}\) with
increasing temperature about \(\phi=0\) will give
more contribution from smaller \(t\). On the other hand, a 
broadening of the distribution for \(u\) in the form 
\(e^{-\beta K u^2/2}\) will lead to more contributions
from t with larger magnitude, since \(t(u)\) increases exponentially in
magnitude when lowering \(u\).

The weak temperature dependence of the electronic coupling, and hence
of the band gap from eq. (\ref{gap}), 
are in strong contrast to the experiment by Porath and co-workers.
It also calls to question the earlier analysis by 
Y. Berlin {\it et al} \cite{ratner} who tried to explain the 
temperature dependence of the observed voltage gap \(U_c\)  
from a reduction of the bandwidth by only including torsional 
motion of base pairs. Our results imply   
a temperature independent \(U_c\) due to the cancellation of the two modes.
The extremely strong increase 
of the gap with temperature could well be due to other effects\cite{resonant}. 
The smallness of \(t\) does not allow much decrease of the bandwidth anyway.

\section{\label{sec:level4}Estimation of parameters for charge transfer}

Single electron or hole transfer
is likely to be very sensitive to the actual motion of
base pairs. Experimental evidence comes from the observation
of two different time scales (5 and 75 ps) 
in charge transfer experiments\cite{2timescales}. The longer time
stems presumably from a necessary reorientation of base pairs in order
to make charge transfer possible\cite{2timescales,bruinsma}. 
Theoretically, torsional acoustic modes 
are indeed found to be extremely soft (\(\leq 20 \text{cm}^{-1}\)) \cite{soft}
and hence slow, which is consistent with dynamic Stoke shifts in
fluorescence spectra\cite{stoke}.
According to figure \ref{fig:fig7},
the magnitude of the transfer integral oscillates between
approximately 0 and 0.1-0.2 eV when one
assumes a standard deviation of the twist angle of about
\(10^o\) (actually more like \(7.5^o\)).
Under this assumption electron tunneling is possible on a
time scale of the  oscillation time \(T_S\)
which can be estimated as \(T_S=2\pi \sqrt{I/S}\),
where \(S\) is the shear
constant and \(I=I_1I_2(I_1+I_2)^{-1}\) the reduced moment
of inertia of two base pairs.

More specifically, single charge transfer rates between two
base pairs can be estimated from the
classical Marcus formula \cite{marcus} (neglecting nuclear
tunneling effects)
\begin{equation}
k_{CT}=\nu_n\,\kappa_{el}\,e^{-\beta\Delta E^*},\label{eq:marcus}
\end{equation}
where \(\nu_n\) and \(\Delta E^*=(E_\lambda+E_o)^2/(4E_\lambda)\)
are the effective nuclear frequency and energy
to reach the transition state, while \(\kappa_{el}\) describes
the electronic
transmission coefficient. \(E_\lambda\) is the total
reorganization energy and \(E_o\le 0\) is the
reaction energy which is zero for charge transfer between
homogeneous base pairs.

We expect the charge transfer between 
neighboring base pairs to be adiabatic, i.e.
\(\kappa_{el}\approx 1\), since the Landau-Zener
probability P for crossing the surfaces at the transition state 
is close to unity with \cite{LandauZener} 
\begin{eqnarray}
  P&=&1-exp(-2\pi\gamma)\label{eq:gamma1}\\
  \text{and}\qquad\gamma&=&\frac{t^2}{8 \hbar \nu_n \sqrt{\pi k_B T \Delta E^*}}.\label{eq:gamma2}
\end{eqnarray}
This is because \(\gamma\) is large due to the relatively large transfer integral t. 
Inner shell reorganization energies and nuclear frequencies associated with 
adjusting
the bonds between atoms of a single base pair can be calculated
by the method of A. Klimk\=ans and S. Larsson described in \cite{benzine}.
The results are shown in table \ref{tab:table2}, which also
contains estimates for moments of inertia.
The frequencies
are quite similar to in-plane \(E_{2g}\) frequencies
(\(4.8\,10^{13} Hz\)) in graphite \cite{graphite},
and our calculated inner shell reorganization energies lie
between those of benzene (0.26 eV) and anthracene
(0.07 eV) \cite{benzine}, as one would expect from the number
of aromatic rings involved. Since
the total reorganization energy is estimated from experiments to be
only about 0.4 eV \cite{outerreorg} which is close
to our values for the inner reorganization energies from
table \ref{tab:table3}, it suggests that the base pairs
must be well protected by the double helix formation and
solvation effects are minor. The small
reorganization energies also contribute to
the adiabaticity by increasing \(\gamma\) in equation \ref{eq:gamma2}.

\begin{table}
\caption{\label{tab:table2} Inner reorganization energies \(E_{\lambda,i}\), moments of inertia \(I\) and nuclear frequencies \(\nu_n\) of single base pairs. The
calculations are done with \textsc{SIESTA} and DZP basis set. \(E_{\lambda,i}\) is determined by the energy difference of a hole/electron doped base pair between doped and undoped relaxed geometries. The moment of inertia is estimated with \(I=\sum m_ir_i^2\) where the sum is over all atoms of a base pair and \(r_i\) are measured from the helical axis of B-DNA. The nuclear frequencies are derived with \(E_{\lambda,i}=2(\pi\nu)^2\sum \mu_{i,i+1} u_{i,i+1}^2\) with \(\mu_{i,i+1}\) and \(u_{i,i+1}\) being the reduced mass and change in bond length of atoms i and neighbor i+1 when comparing doped with undoped relaxed structures.   }
\begin{ruledtabular}
\begin{tabular}{cc|cc}
                      &            &   G\(\cdot\)C &   A\(\cdot\)T        \\ \hline
\(E_{\lambda,i}\) [eV]&   hole     &     0.18    &      0.08          \\ \cline{2-4}  
                      &  electron  &     0.28    &      0.20          \\ \hline
\(I[10^{-44}\text{kg\,m}{}^2]\)&   &     5.0       &      6.2             \\ \hline
\(\nu_n[10^{13}1/\text{s}]\)& hole &     3.97    &      3.64          \\ \cline{2-4}
                      &   electron &     5.09    &      2.74          
\end{tabular}
\end{ruledtabular}
\end{table}

\begin{table}
\caption{\label{tab:table3} Oscillation time of twisting motion \(T_S\), total reorganization energy 
\(E_\lambda\approx\sum E_{\lambda,i}\), Franck-Condon factor FC, and upper/lower limits of adiabatic 
charge transfer rates between two neighboring
 base pairs (bps). The rates are calculated with eq.(\ref{eq:marcus}) and data from table (\ref{tab:table2})
in the normal regime. 
The upper limit corresponds to a nuclear frequency \(1/\nu=1/\nu_n^1+1/\nu_n^2\), the lower limit to \(\nu=1/T_S\).}
\begin{ruledtabular}
\begin{tabular}{cc|ccc}
     sequence(ref. \cite{f3})
                                 &            &      GG       &      AG(ref. \cite{f4})
                                                                              &      AA       \\ \hline\hline
     \(T_S\)[ps]                 &            &    2.02       &     2.13      &     2.25      \\ \hline
 \(E_\lambda\)[eV]               &   hole     &    0.36       &     0.26      &     0.16      \\ \cline{2-5}
                                 &  electron  &    0.57       &     0.48      &     0.39      \\ \hline
     FC (at 300K)                &   hole     &    0.03       &     0.88      &     0.22      \\ \cline{2-5}
                                 &  electron  &    0.004      &     0.01      &     0.02      \\ \hline\hline
                                 &   limit    &               &               &               \\ \hline
\(k_{hole}\) \([1/s]\)           &  upper     &\(6.0\,10^{11}\)&\(1.7\,10^{13}\)&\(4.1\,10^{12}\) \\ \cline{2-5}  
                                 &  lower     &\(1.5\,10^{10}\)&\(4.1\,10^{11}\)&\(9.9\,10^{10}\) \\ \hline
\(k_{electron}\)\([1/s]\)        &  upper     &\(1.1\,10^{11}\)&\(1.8\,10^{11}\)&\(3.1\,10^{11}\) \\ \cline{2-5}
                                 &  lower     &\(2.1\,10^9\)   &\(4.7\,10^{9}\)&\(1.0\,10^{10}\)  
\end{tabular}
\end{ruledtabular}
\end{table}

Upper and lower bounds for the charge transfer rates between
two base pairs are shown in table \ref{tab:table3}.
For the upper bound, we take \(\nu\) in equation (\ref{eq:marcus})
to be the inner base pair nuclear frequency,
while for a lower bound we take it to be 
the base pair twist frequency. This slow process can be 
viewed as reaching an encounter complex with a frequency \(\nu=1/T_S\).
We note the large Franck-Condon factor in
table \ref{tab:table3} for hole transfer between 
\(A\cdot T\) and \(G\cdot C\) (dimer AG). 
This is because this case is 
close to the so-called optimal electron
transfer regime in which the reorganization energy
(\(0.26\) eV) is largely compensated by the
reaction energy (\(0.20\pm0.05\) eV) \cite{Eo} as estimated from 
analyzing electron transfer yield data.
Note that this reaction energy is smaller than the difference of the bare
oxidation potentials (\(0.4 - 0.5\) eV) between
isolated adenine and guanine bases in a polar solvent \cite{oxpot_solvent},
in gas phase \cite{oxpot_gas}, or from
{\it ab initio} calculations and Koopman's theorem\cite{band_IP,Becky}.
Furthermore, hole transfer is in most cases at least
one order of magnitude faster and hence more efficient than
electron transfer. 

In order to compare with experiment we turn
to (hole) transfer between initially
excited 2-aminopurine (an isomer of adenine) and
guanine \cite{barton}. The measured charge injection time of
10ps (corresponding to a rate of \(10^{11}1/s\)) is
surprisingly close to our lower bound value
\(4.1\,10^{11}1/s\). Hence we speculate that the
twisting motion of base pairs is rate limiting.

Another simple application of the 2 base pair
charge transfer rates of table \ref{tab:table3} is the hole
hopping between homogeneous sequences, e.g. \((A\cdot T)_N\),
when one assumes (biased) random walk
as the underlying 
mechanism of the charge migration. In this case, the rate
shows a weak algebraic
distance dependence\cite{jortner}
\(k_{CT}\approx k_{\text{HOP}}N^{-\eta}\), where N is the
number intervening
base pairs and \(1\le\eta\le2\). Absolute rates can be
predicted with our (lower bound) value
\(k_{\text{HOP}}=9.9\, 10^{10}1/s\) \(\approx10^{11}1/s\).
In order to compare with
experiment, e.g. \cite{lewis}, one would have to include the initial
transfer from the donor to the bridge,
as well as the final transfer from the bridge to the acceptor
which leads to a further reduction of the rate. 
General formulas are provided in refs. \cite{Eo,jortner}. 

Finally, since charge transfer between adjacent base pairs is likely to be
adiabatic, one should avoid including these when fitting to a diabatic
charge transfer rate \(k_{CT}\sim \exp(-\beta R)\).
Including the adjacent base pair charge transfer in the fit 
leads to an under-estimation of \(\beta\).

\section{\label{sec:level5}Conclusions}

In summary, we have studied the electronic properties of
DNA from several perspectives.
Our initial motivation was to come up with possible
conditions which enhance the conductance of DNA. For example,
we expected a stretched ribbon-like DNA
to be a potential candidate for a molecular wire. Unfortunately,
that has not turned out
to be the case. Indeed, in the ribbon-like DNA,
the base pair separation can become quite small.
However, the pp\(\pi\) interaction appears not
very effective as there are only very few good contacts. 
Using the \textit{ab initio} code SIESTA we compared
transfer integrals between
base pairs of A, B and stretched DNA. Our calculations
suggest that A-DNA and stretched, ribbon-like DNA should
support electrical current even less than B-DNA.

On the other hand, a first principle study of 4 base pair long DNA in the B-form
with different counterions (sodium, magnesium) under 
dry and wet (\(\sim 200\) water molecules) conditions
shows the possibility of small activation gaps of order \(k_BT\). 
Although the \(\pi\) -  \(\pi^*\) gap is several eV large, a
small gap is formed mainly by water and counterion states in the case of wet DNA,
or by backbone and counterion states in the case of dry DNA.
In the latter case, the backbone states rise in energy and lie higher
than the occupied \(\pi\) MOs 
due to repulsive interactions of the negatively charged backbone.
Since small activation gaps (\(\sim 0.1 eV\)) are found experimentally
for bundles\cite{kawai,rakitin} or supercoiled DNA \cite{fet,tran} 
from a sodium buffer, this could be caused
by residual water and condensed counterions providing an alternative
pathway for charge hopping.
To our knowledge, there is less experimental 
conductivity data available for DNA from a magnesium buffer. 
The most prominent case is the 
observed induced superconductivity \cite{proximity} requiring truly extended
metallic-like states. From our first principle studies 
of DNA with ions and water, we do notice some differences between 
magnesium and sodium ions. For wet DNA, 
electron excitation from occupied water and magnesium states 
into empty \(\pi^*\) states seems possible making
electron conduction feasible. In addition, for dry DNA
with magnesium ions, the activation gap between occupied backbone
and unoccupied magnesium states is vanishingly small. 
In addition, there is a large DOS of
phosphate states right below the Fermi energy available.
This pinning-effect opens the possibility of a pathway 
for hole hopping through the backbone. 
 
In order to get a better microscopic understanding
of the electronic coupling between adjacent base pairs 
and how it is affected by
twisting and displacement motion, we developed
a H\"uckel-Slater-Koster model with parameters
obtained from \textit{ab initio} calculations. We find
that the transfer integral is constrained by a
competition between 
pp\(\sigma\) and pp\(\pi\) interaction and interference
by phase factors of atomic \(p_z\) orbitals.
The latter effect also leads
to sign changes in the transfer integrals as a function
of the torsional angle not anticipated in previous studies.

Furthermore, the temperature dependence of the average
transfer integral and band gap is expected to be weak:
First, the electronic coupling is already weak anyway,
and cannot be reduced much.
Second, twisting and displacement 
motion of base pairs have cancelling contributions.

Next, we showed that charge transfer rates between adjacent
base pairs are likely in the adiabatic limit. 
Thus, if one uses (modified) bases as donor and acceptor 
in charge transfer experiments, it is important to take this into account. 
When obtaining the decay length \(\beta\) of diabatic charge transfer
from a fit to an experimental rate as a function of distance, 
the data point corresponding to the 
donor and acceptor next to each other should not be included in the fit. 

We also calculated Franck-Condon factors,
reorganization energies and nuclear frequencies
to obtain absolute
rate estimates. Furthermore, we suggest that the
twisting motion might be rate limiting for hole transfer.
Due to the adiabaticity of the two base pair rates, 
DNA might be a good medium for diffusive charge transport.\\

\textit{Acknowledgements}. We would like to thank P. Ordej\'on,
E. Artacho, D. S\'anchez-Portal and J. M. Soler
for providing us with their \textit{ab initio} code SIESTA, 
as well as S. Dixit for providing us with the Dickerson dodecamer
structure from classical molecular dynamics.
We acknowledge helpful conversations with J.K. Barton, C.Y. Fong,
and J. Jortner.  This work was supported by the U.S. Department of Energy,
Office of Basic Energy Sciences, Division of Materials Research,
by a faculty seed grant from the U.C. Davis office of research, by
a collaborative seed grant from the Materials Research Institute of
Lawrence Livermore National Laboratories, and by the National Science
Foundation grant DMR-9986948.

\appendix
\section{\label{sec:levelA}Appendix}
In the following we derive a general formula for the calculation of interaction matrix elements 
between two atomic \(p_z\) orbitals 
belonging to adjacent base pairs.  
In general, the orbitals point in different directions, since the local z-axis is different for each base 
mainly due to a propeller twist angle, in case of A-DNA also because of a large inclination angle.
We first expand the orbitals \(p_z, p_z'\) in a global coordinate system with the z-axis pointing in the direction of 
the helix   
\begin{eqnarray}
  p_z&=&\alpha\tilde p_x+\beta\tilde p_y + \gamma\tilde p_z\label{eq:pz}\\
  p_z'&=&\alpha'\tilde p_x+\beta'\tilde p_y + \gamma'\tilde p_z\nonumber,
\end{eqnarray}
where the expansion coefficients are the direction cosines. If \((l,m,n)\) is the unit vector pointing from one 
orbital to the other, the Slater-Koster relations \cite{SK} 
\begin{eqnarray}
  E_{xx}&=&l^2V_{pp\sigma}+(1-l^2)V_{pp\pi}\label{eq:E}\\
  E_{yy}&=&m^2V_{pp\sigma}+(1-m^2)V_{pp\pi}\nonumber\\
  E_{zz}&=&n^2V_{pp\sigma}+(1-n^2)V_{pp\pi}\nonumber\\
  E_{xy}&=&lm(V_{pp\sigma}-V_{pp\pi})\nonumber\\
  E_{xz}&=&ln(V_{pp\sigma}-V_{pp\pi})\nonumber\\
  E_{yz}&=&mn(V_{pp\sigma}-V_{pp\pi})\nonumber
\end{eqnarray}
can be used to obtain the following formula for the interaction matrix element 
\begin{eqnarray}
  V&=&\alpha\alpha'E_{xx}+\beta\beta'E_{yy}+\gamma\gamma'E_{zz}\nonumber\\
    &&+(\alpha\beta'+\beta\alpha')E_{xy}+(\alpha\gamma'+\gamma\alpha')E_{xz}\nonumber\\
    &&+(\beta\gamma'+\gamma\beta')E_{yz}\label{eq:VSK}.
\end{eqnarray}
The Slater-Koster matrix elements \(V_{pp\sigma}\) and \(V_{pp\pi}\) are given by eq. (\ref{eq:ppX}) and contain the distance dependence.


\begin{references}

\bibitem{1962} D. D. Eley and D. I. Splivey, Trans. Faraday Soc., {\bf 58},  411 (1962)
\bibitem{young} M. A. Young, G. Ravishanker and D. L. Beveridge, Biophys. J., {\bf 73}, 2313 (1997)
\bibitem{adna} P. J. de Pablo {\it et al}, Phys. Rev. Lett. {\bf 85}, 4992 (2000)
\bibitem{porath} D. Porath, A. Bezryadin, S.  De Vries, and C. Decker, Nature (London) {\bf 403}, 635 (2000)
\bibitem{fink} H. W. Fink and C. Sch\"{o}nenberger, Nature {\bf 398},
407 (1999).
\bibitem{proximity} A. Y. Kasumov {\it et al}, Science {\bf 291}, 280 (2001)
\bibitem{kawai} L. Cai, H. Tabata, and T. Kawai, Appl. Phys. Lett. {\bf 77}, 3105 (2000)
\bibitem{fet} K. -H. Yoo {\it et al}, Phys. Rev. Lett. {\bf 87}, 1981 (2001)
\bibitem{siesta} D. S\'anchez-Portal, P. Ordej\'on, E. Artacho,  and J. M. Soler, Int. J. Quantum Chem. {\bf 65}, 453 (1997);
                 E. Artacho, D. S\'anchez-Portal, P. Ordej\'on, A. Garcia, and J. M. Soler, Phys. Status Solidi (b) {\bf 215}, 809 (1999); P. Ordej\'on, E. Artacho, J. M. Soler, Phys. Rev. B {\bf 53}, R10441 (1996)
\bibitem{SK} J. C. Slater, G. F. Koster, Phys. Rev. {\bf 94}, 1498 (1954)
\bibitem{harris} W. A. Harrison, Electronic Structure and the Properties of Solids, Dover Publications, 
                 Inc., New York, p.48,481 (1989)
\bibitem{barton} C. Wan, T. Fiebrig, O. Schiemann, J.  K.  Barton, A.  H. Zewail, Proc. Natl. Acad. Sci. {\bf 97}, 14052 (2000)
\bibitem{Bform} R. Chandrasekaran and S. Arnott, J. Biomol. Struct.
Dynamics, {\bf 13}, 1015 (1996)
\bibitem{Aform} S. Arnott and D. W. L. Hukins, Biochem. Biophys. Res. Commun., {\bf 47}, 1504 (1972)
\bibitem{comb} D. Bensimon, A. J. Simon, V. Croquette, A. Bensimon, Phys. Rev. Lett. {\bf 74}, 4754 (1995);
               J. F. Allemand {\it el al}, Biophys. J., {\bf 73}, 2064 (1997);
               I. Parra and B. Windle, Nature Genetics, {\bf 5}, 17 (1993);
               C. Bustamante, S. B. Smith, J. Liphardt, and D. Smith, Curr. Opin. Struct. Biol. {\bf 10}, 279 (2000)
\bibitem{MD} A. Lebrun and R. Lavery, Nucl. Acids Res., {\bf 24}, 2260 (1996);
             M. W. Konrad and J. I. Bolonick, J. Am. Chem. Soc., {\bf 118}, 10989 (1996);
             K. M. Kosikov {\it et al}, J. Mol. Biol., {\bf 289}, 1301 (1999)
\bibitem{ps} N. Troullier, J. L. Martins, Phys. Rev. B {\bf 43}, 1993 (1991)
\bibitem{klein} L. Kleinman, D. M. Bylander, Phys. Rev. Lett. {\bf 48}, 1425 (1982)
\bibitem{sankey} O. F. Sankey, D. J. Niklewski, Phys. Rev. B {\bf 40}, 3979 (1989)
\bibitem{gga} J. P. Perdew, K. Burke, M. Ernzerhof, Phys. Rev. Lett. {\bf 77}, 3865 (1996)
\bibitem{CH3} terminal CH\({}_3\) groups replace the backbone
\bibitem{newton} M. D. Newton, J. Phys. Chem., {\bf 90}, 3734 (1986)
\bibitem{f1} Corresponds to perpendicular base pair separation of A-DNA which has a tilt angle of \(20^o\). Note the huge difference to A-form GG.
\bibitem{f2} Perfectly parallel aligned base pairs. Note the sign change compared to B-form GG at \(36^o\).
\bibitem{band_IP} H. Sugiyama and I. Saito, J. Am. Chem. Soc. {\bf 118}, 7063 (1996)
\bibitem{band} M. L. Zhang, M. S. Miao, V. E. van Doren, J. J. Ladik, J. W. J. Mintmire, J. Chem. Phys. {\bf 111}, 8696 (1999)
\bibitem{elcoupl}A. A. Voityuk, J. Jortner, M. Bixon, and N. R\"osch, J. Chem. Phys. {\bf 114}, 5614 (2001)
\bibitem{fit} http://www.ulib.org/webRoot/Books/Numerical\underline{ }Recipes/\\bookc.html
\bibitem{tran} P. Tran, B. Alavi, and G. Gruner, Phys. Rev. Lett. {\bf 85}, 1564 (2000)
\bibitem{hopfinger} A. J. Hopfinger, Intermolecular interactionns and bimolecular organization 
                    (Wiley, New York, 1977) ch.7
\bibitem{Cai2} L. Cai, H. Tabata and T. Kawai, Nanotechnology {\bf 12}, 211 (2001)
\bibitem{storm} A. J. Storm {\it et al}, Appl. Phys. Lett., {\bf 79}, 3881 (2001)
\bibitem{rakitin} A. Rakitin {\it et al},  Phys. Rev. Lett. {\bf 86}, 3670 (2001)
\bibitem{ions} M. A. Young, B. Jayaram, and D. L. Beveridge, J. Am. Chem. Soc. {\bf 119}, 59 (1997);
               A. P. Lyubartsev, A. Laaksonen, J. Biomol. Struct. Dynamics {\bf 16}, 579 (1998);
               Alexandre M. J. J. Bonvin, Eur. Biophys. J. {\bf 29}, 57 (2000);
               R. N. Barnett {\it et al}, Science {\bf 294}, 567 (2001)
\bibitem{mulliken} R. S. Mulliken, J. Chem. Phys. {\bf 23}, 1833 (1955)
\bibitem{Fermi} To facilitate convergence, SIESTA can employ a finite
temperature in the occupancy functions, which allows for a unique
determination of the Fermi energy; for $T=0K$ this is of course
indeterminate, and we have taken the Fermi energy at the midpoint of the
HOMO-LUMO gap.  
\bibitem{solv_ions} M. C. Vicens and G. E. L\'opez, J. Comp. Chem. {\bf 21}, 64 (2000)
\bibitem{huckel} S. Priyadarshy, S. M. Risser and D. N. Beratan, J. Phys. Chem. {\bf 100}, 17678 (1996)
\bibitem{ratner} Y. A. Berlin, A. L. Burin, and M. A. Ratner, Superlattices Microstruct. {\bf 28}, 241 (2000)
\bibitem{resonant} M. Hjort and S. Strafstr\"om, Phys. Rev. Lett. {\bf 87}, 228101/1-4 (2001)
\bibitem{2timescales} C.  Wan {\it et al}, Proc. Natl. Acad. Sci. USA{\bf 96}, 6014 (1999)
\bibitem{bruinsma} R. Bruinsma, G. Gruner, M. R. D'Orsogna, J. Rudnick, 
                   Phys. Rev. Lett. {\bf 85}, 4393 (2000)
\bibitem{soft} S. Cocco and R. Monasson, J. Chem. Phys. {\bf 112}, 10017 (2000)
\bibitem{stoke} E. Brauns {\it et al}, J. Am. Chem. Soc. {\bf 121}, 11644 (1999)
\bibitem{marcus} R. A. Marcus and N. Sutin, Biochim. Biophys. Acta
{\bf 811}, 265 (1985)
\bibitem{LandauZener} L. D. Landau, Phys. Z.  Sowjetunion {\bf 1}, 88 (1932); {\bf 2} 46 (1933);
                      C. Zener, Proc. R. Soc. London A {\bf 137}, 696 (1932); {\bf 140} 660 (1933)
\bibitem{benzine} A. Klimk\=ans and S. Larsson, Chem. Phys. {\bf 189}, 25 (1994)
\bibitem{graphite} C. T. Chan, K. M. Ho and W. A. Kamitakahara, Phys. Rev. B {\bf 36}, 3499 (1987)
\bibitem{outerreorg} A. Harriman, Angew. Chem. Int. Ed. Engl. {\bf 38}, 945 (1999)
\bibitem{f3} Electron transfer actually involves the complementary bases C and T.
\bibitem{f4} We used the following reaction energies: hole transfer between A and G, \(E_o=-0.2\) eV; electron transfer between T and C, \(E_o\approx 0\) eV.
\bibitem{Eo} M. Bixon and J. Jortner, J. Am. Chem. Soc. {\bf 123}, 12556 (2001)
\bibitem{oxpot_solvent}  C. A. M. Seidel, A. Schulz and M. H. M. Sauer, J. Phys. Chem. {\bf 100}, 5541 (1996);
                         S. Steenken and S. V. Jovanovic, J. Am. Chem. Soc. {\bf 119}, 617 (1997)
\bibitem{oxpot_gas} V. M. Orlov, A. N. Smirnov and T. M. Varshavsky, Tet. Lett. {\bf 48}, 4377 (1976);
                 S. G. Lias, J. E. Bartmess, J. F. Liebman, L. J. Holmes, 
                 R. D. Levin, W. G. Mallard, J. Phys. Chem. Ref. Data 17 (1988)
\bibitem{Becky} S. D. Wetmore, R. J. Boyed and L. A. Eriksson, Chem. Phys. Lett. {\bf 322}, 129 (2000)
\bibitem{jortner} J. Jortner, M. Bixon, T. Langenbacher and M. E. Michel-Beyerle, Proc.  Natl.  Acad. Sci. USA {\bf 95}, 12759 (1998) 
\bibitem{lewis} F. D. Lewis {\it et al}, Science {\bf 277}, 673 (1997)

\end{references}
\end{document}